\documentclass[a4paper,11pt]{article}
\pdfoutput=1 

\usepackage{jcappub} 
                     
\usepackage{siunitx}

\usepackage[T1]{fontenc} 

\title{Isotropy statistics of CMB hot and cold spots}

\author[a]{Md Ishaque Khan,}
\author[a,1]{Rajib Saha \note{Corresponding author.}}

\affiliation[a]{Department of Physics, Indian Institute of Science Education and Research, \\Bhopal-462066, India}

\emailAdd{ishaque16@iiserb.ac.in}
\emailAdd{rajib@iiserb.ac.in}

\abstract{Statistical Isotropy of the Cosmic Microwave Background (CMB) radiation has been studied and debated extensively in recent years. Under this assumption, the hot spots and cold spots of the CMB are expected to be uniformly distributed over a 2-sphere. We use the orientation matrix, first proposed by Watson (1965) and Scheidegger (1965) and associated shape and strength parameters (Woodcock, 1977) to analyse whether the hot and cold spots of the observed CMB temperature anisotropy field are uniformly placed. We demonstrate the usefulness of our estimators by using simulated toy models containing non-uniform data. We apply our method on several foreground minimized CMB maps observed by WMAP and Planck over large angular scales. The shape and strength parameters constrain geometric features of possible deviations from uniformity (isotropy) and the power of the anomalous signal. We find that distributions of hot or cold spots in cleaned maps show no unusual signature of clustering or girdling. Instead, we notice a strikingly uniform distribution of hot spots over the full sky. The signal remains robust with respect to the four cleaned maps used and presence or absence of the non-Gaussian cold spot (NGCS). On the partial sky with WMAP KQ75 and Planck U73 masks we find anomalously low strength of non-uniformity for cold spots which is found to be robust with respect to various cleaning methods, masks applied, instruments, frequencies, and the presence or absence of the NGCS. Interestingly we find that the signal of anomalously weak non-uniformity could be due to contributions from the quadrupole and octupole and may be related with the low CMB temperature variance anomaly.}

\begin{document}
\maketitle
\flushbottom

\section{Introduction}
The anisotropies of the CMB temperature field play a fundamental role behind the formation of large scale structure of the universe. Over large angular scales these anisotropies are believed to have originated dominantly due to quantum fluctuations of the inflaton field present in the very early universe. The primordial curvature perturbation power spectrum contains snapshots of these fluctuations at the time of horizon crossing. Under the assumption of rotational invariance of this power spectrum, the observable CMB is expected to be statistically isotropic without picking up any preferred direction in the radiation field. However, in recent literature there are discussions \citep{hajian_statistical_2005,souradeep_hajian_2004, Eriksen_2004, rassat_preferred_2013, Kim_2010, Aluri_2017, Kim_20102, 10.1111/j.1365-2966.2011.19981.x, PhysRevD.70.043515} and claims of the presence of preferred direction \citep{PhysRevD.69.063516, doi:10.1142/S0218271804005948, 10.1111/j.1365-2966.2009.14728.x, PhysRevLett.93.221301, 10.1111/j.1365-2966.2008.12960.x, PhysRevLett.95.071301, PhysRevD.89.023010} in the CMB field or breakdown of rotational invariance of the primordial power spectrum \citep{10.1111/j.1365-2966.2004.08229.x, akrami_power_2014, hansen_power_2009}. These indicate possible hints towards new fundamental physics \citep{PhysRevD.75.083502, MALEKNEJAD2013161, soda_statistical_2012, 10.1093/mnras/sty1689, li_finslerian_2017, dimastrogiovanni_non-gaussianity_2010}. Apart from the cosmological origin of any possible anisotropic signal in the CMB, a question of equal importance is whether the foreground minimized CMB maps may have any residual systematics, e.g., residual foreground contamination which may potentially induce a breakdown from rotational invariance of the field and consequently give rise to a preferred direction. Needless to mention, if the presence of such residuals are not taken care of, cosmological parameters estimated from the CMB maps will be potentially biased or contain inaccurate confidence intervals on their estimated values. Thus, for proper and accurate extraction of cosmological information from CMB maps it is utmost essential to analyse from as many different perspectives as possible, if the observed maps contain any signal of breakdown of statistical isotropy.

In this article, we use a new method to investigate the isotropy of the anisotropy pattern of the CMB radiation field. The local maxima (hot spots) and minima (cold spots) of the CMB are uniformly distributed over the surface of a two-sphere if the field is isotropic and does not contain any preferred direction. A breakdown from uniformity of the distributions of either or both of these types of spots then needs to be investigated to validate the null hypothesis of isotropy. We use the concept of the orientation matrix introduced by Watson (1965) \citep{10.2307.2333824}  and Scheidegger (1965) \citep{scheidegger_statistics_1965}, to probe any violations of uniformity of the distributions of hot and cold spots over the CMB sky. The method is unique in its nature, since if there is any non-uniformity present in the data, it provides a geometric description of the non-uniformity in terms of clustering or girdling (ring structure) and an additional measure of the magnitude of such a deviation from uniformity in terms of the so-called strength parameter.

In earlier literature, the distribution of hot and cold spots of the CMB maps have been shown to encapsulate topological properties of the temperature field \citep{PhysRevD.64.083003}. Apart from possible primordial or residual systematic effects present in the cleaned CMB maps, any unusual features such as clustered or girdled spots in cleaned maps may indicate presence of organised collections of structures or voids \citep{10.1093/mnras/stv488} between the source and observer. In addition, a higher signal-to-noise ratio at the hot spots \citep{10.1093/mnras/216.1.25P_sazhin,10.1093/mnras/226.3.655} as noted by \citep{Larson_2004}, makes them favourable to study.

A statistical analysis of such hot and cold spots on the pixelated sphere using the orientation matrix has some other practical advantages as well. This reduces the large data set of numerous hot and cold spots to their respective eigenbases \citep{jolliffe_principal_2016}. For accurate estimation of properties of the CMB sky, sometimes one analyses partial sky CMB maps, which are obtained by masking out the foreground dominated regions near the plane of the Milky Way \citep{10.1093/ptep/ptu065}. Our study can easily be applied to partial CMB skies, without bothering about the complications that may arise due to multipole mode couplings on masked CMB maps \citep{Hivon_2002} or due to any subtle biases introduced due to inpainting methods \citep{rassat:cea-01135418} when an underlying statistically isotropic CMB theoretical angular power spectrum is assumed while trying to reconstruct the lost sky region due to masking.

Previously, authors of \cite{Larson_2004} have studied one-point statistics such as the number, mean and variance of local extrema and demonstrated that on an average the hot (cold) spots of the observed CMB data are not hot (cold) enough. Further with the help of two-point statistics it was shown that possibly unusual properties associated with large angular scale structures are related with the behaviour of hot and cold spots of the CMB \citep{ 10.1111/j.1365-2966.2009.14810.x}. Since the initial density fluctuations are assumed to be Gaussian, a 3D Gaussian field model \citep{1986ApJ...304...15B} was extended to address the 2D Gaussian CMB temperature field \citep{10.1093/mnras/226.3.655}. Notable works that have followed thereafter have dealt with the number density, shapes, separation distances and peak-peak correlation functions for spots \citep{PhysRevD.84.083510,10.1111/j.1365-2966.2011.18563.x, 10.1093/mnras/stab368}. The discovery of the non-Gaussian cold spot \citep{Vielva_2004,10.1111/j.1365-2966.2004.08419.x,Cruz_2007} and the theoretical frameworks for hot and cold spots of a 2D Gaussian field as presented in \citep{10.1093/mnras/226.3.655, PhysRevD.84.083510, 10.1111/j.1365-2966.2011.18563.x} have inspired further research to probe non-Gaussianity \citep{10.1093/mnras/stab368, Chingangbam_2012, 10.1111/j.1365-2966.2009.15503.x}.

Our paper is organised as follows. In Section \ref{estim} we define the estimators for the shape and strength of non-uniformity used in our study. In Section \ref{toy} we analyse the behaviour of the chosen estimators on two toy model maps to illustrate clustering or girdling of spots. In Section \ref{analysis_data}, we discuss the application of these estimators to observed CMB maps. In Section \ref{results} we present our results for any non-uniformity of spots in the observed CMB. In Section \ref{quadoct}, we assess if the results obtained may be related with low CMB variance.  In Section \ref{h+c} we consider the composite set of both hot and cold spots and assess their uniformity. In Section \ref{conc} we form general conclusions and summarise our findings.

\section{Estimators}\label{estim}
One can study the distribution of spherical data as an analogue of unit mass points on the surface of a 2-sphere with the help of a unit-mass orientation matrix \citep{10.2307/30060119, WOODCOCK1983539, fisher_lewis_embleton_1987},

\begin{equation}\label{orien1}
    T=\left(
\begin{array}{ccc}
   \sum_i{x_i}^2  & \sum_i x_i y_i & \sum_i x_i z_i \\
   \sum_i x_i y_i & \sum_i {y_i}^2 & \sum_i y_i z_i\\
    \sum_i z_i x_i & \sum_i z_i y_i & \sum_i {z_i}^2
\end{array}
\right),
\end{equation} where, $(x_1,y_1,z_1)...(x_n,y_n,z_n)$ are the direction cosines of unit mass points labelled with index $i=1,...,n$. Scheidegger (1965) \citep{scheidegger_statistics_1965} used the principal eigenvector of the normalised unit-mass orientation matrix (i.e, $T/n$) to find a `mean' axis, while Watson (1965) \citep{10.2307.2333824} used the eigenvalues of $T$ to classify non-uniform placements of data points on a 2-sphere. Woodcock (1977) \citep{10.1130/0016-7606(1977)88<1231:SOFSUA>2.0.CO;2} defined two kinds of eigenvalue ratios to quantify the shape and strength of such non-uniformity.

The CMB has a nearly uniform background temperature of $T_0=2.726K$ \citep{Fixsen_2009}. However there exist small directionally dependent differences of the order of a few hundred $\mu K$ relative to $T_0$, which are called anisotropies. The hot and cold spots of the CMB temperature anisotropy field can therefore be treated as data points on a 2-sphere and their placements can be studied with the help of the orientation matrix.

In \citep{fisher_lewis_embleton_1987}, the authors treat all data points on the sphere as equivalent in magnitude and ascribe unit masses to the same. However, in the case of the CMB temperature field, as various extrema are offset differently relative to $T_0$, we extend the concept to include `non-unit masses', i.e, peak values of the hot spots or cold spots, and express an orientation matrix $\mathcal{T}^{(s)}$ as,

\begin{equation}\label{orien_matrix}
    \mathcal{T}^{(s)}=\frac{\sum_i m^{(s)}_i \tau^{(s)}_i}{\sum_i m^{(s)}_i}
\end{equation} where,
\begin{equation}
    \tau^{(s)}_i=\left(
\begin{array}{ccc}
   {x^{(s)}_i}^2  & x^{(s)}_i y^{(s)}_i & x^{(s)}_i z^{(s)}_i \\
   x^{(s)}_i y^{(s)}_i & {y^{(s)}_i}^2 & y^{(s)}_i z^{(s)}_i\\
    z^{(s)}_i x^{(s)}_i & z^{(s)}_i y^{(s)}_i & {z^{(s)}_i}^2
\end{array}
\right).
\end{equation} Here, the superscript ${}^{(s)}$ stands for $s=h,c$, for hot or cold spots, respectively. For the $i^{th}$ s-spot, $x_i^{(s)},y_i^{(s)},z_i^{(s)}$ are its direction cosines. Its non-unit mass weight is $m^{(s)}_i=|\Delta T^{(s)}_i|=|T^{(s)}_i-T_0|$, which is the magnitude of its temperature relative to $T_0$. The normalisation by the non-unit mass weights in equation \eqref{orien_matrix} ensures that the sum of the three eigenvalues of the non-unit mass orientation matrix ($\mathcal{T}^{(s)}$) used by us becomes unity, as it was in the case of the normalised unit-mass orientation matrix ($T/n$) \citep{fisher_lewis_embleton_1987}.

The principal advantage of using non-unit masses is that it helps take into consideration the additional randomness from peak values of the spots along with the randomness that comes from eigenvector directions. In the case of perfect uniformity in the placement of spots, there can be no preferred eigenvector directions, and hence all eigenvalues of the orientation matrix must be equal. Inequalities of eigenvalues therefore indicate the presence of non-uniformity. We note that each spot on the CMB sphere may contribute differently in terms of its mass weights ($|\Delta T_i^{(s)}|$) to determine the magnitudes of eigenvalues and hence the preference of any eigenvector direction. Thus the inclusion of non-unit mass weights is crucial for an accurate detection of non-uniformity in the arrangement of spots.  In Section \ref{toy}, we further elucidate this using two toy maps. 

The orientation matrix is positive definite by construction, and thus all its eigenvalues ($\lambda_i^{(s)}$ for $i=1,2,3$) are positive and its eigenvectors are mutually orthogonal. Considering the three eigenvalues arranged in ascending order, i.e., $0\leq\lambda^{(s)}_1\leq\lambda^{(s)}_2\leq\lambda^{(s)}_3$, one can quantify the manner and extent of non-uniformity in placements of hot and cold spots about their respective eigenvectors. Isotropic or completely uniform distributions of spots on the CMB sphere correspond to $\lambda_1^{(s)}=\lambda_2^{(s)}=\lambda_3^{(s)}$; for planar girdles of spots that are evenly placed in great circles, $\lambda_1^{(s)}<\lambda_2^{(s)}\simeq \lambda_3^{(s)}$; whereas linear clusters manifest as $\lambda_1^{(s)}\simeq\lambda_2^{(s)} <\lambda_3^{(s)}$.  Thus, the following ratios,

\begin{eqnarray}
\gamma^{(s)}&=&\frac{\log{\frac{\lambda^{(s)}_3}{\lambda^{(s)}_2}}}{\log{\frac{\lambda^{(s)}_2}{\lambda^{(s)}_1}}},\nonumber\\
\zeta^{(s)}&=&\log{\frac{\lambda^{(s)}_3}{\lambda^{(s)}_1}},
\end{eqnarray} can help us to study the nature of placement of spots on the CMB sphere. These are known as the shape and strength parameters, respectively. By definition both these parameters take positive values. The shape parameter describes the arrangement of spots on the sphere, in terms of girdling ($\gamma^{(s)}<1$) as opposed to clustering ($\gamma^{(s)}>1$) and transitions ($\gamma^{(s)}\to1$) between these two shapes of arrangement. The strength parameter quantifies the degree of non-uniformity, starting from a value of zero which corresponds to the case when each of the three eigenvalues are equal to $1/3$ and the distribution of spots is absolutely uniform or isotropic. We present the ranges of values of these estimators and associated interpretations in Table \ref{table_estim}. This table may be helpful as a quick reference to categorise the ways in which hot and cold spots are placed on the celestial sphere of the CMB.

\begin{table}
    \centering
    \caption{Value-based interpretations of isotropy estimators: The shape parameter categorises non-uniformly placed spots into clusters or girdles (rings). The strength parameter tells us how weak or strong is the extent of this non-uniformity. Here, $^{(s)}$ can be replaced by $s=h,c$ for hot spots or cold spots, respectively.}
    \label{table_estim}
    \vspace{0.4cm}

\begin{tabular}{|c |c|l|} 
\hline 
Isotropy estimators & Ranges & Interpretations \\ [1ex] 
 \hline
Shape $\gamma^{(s)}$ & $0\leq\gamma^{(s)}\lesssim \infty$ &  Girdling \\
&&$\quad$ for $\gamma\in[0,1)$ ;\\
&&  Clustering \\
&&$\quad$ for $\gamma\in(1,\infty)$ ;\\
&&  Cluster-girdle\\
&&$\quad$ transitions \\
&&$\quad$ for $\gamma^{(s)}\to 1$ .\\[1ex]
Strength $\zeta^{(s)}$ & $0\leq\zeta^{(s)}\lesssim\infty$ & Perfect uniformity\\
&&$\quad$ for $\zeta=0$; \\
&& Perfect non-uniformity\\
&&$\quad$ for $\zeta\to\infty$ .\\ [1ex]
 \hline
\end{tabular}
\end{table}

\begin{figure}[htbp]
\centering
    \includegraphics[width=\textwidth]{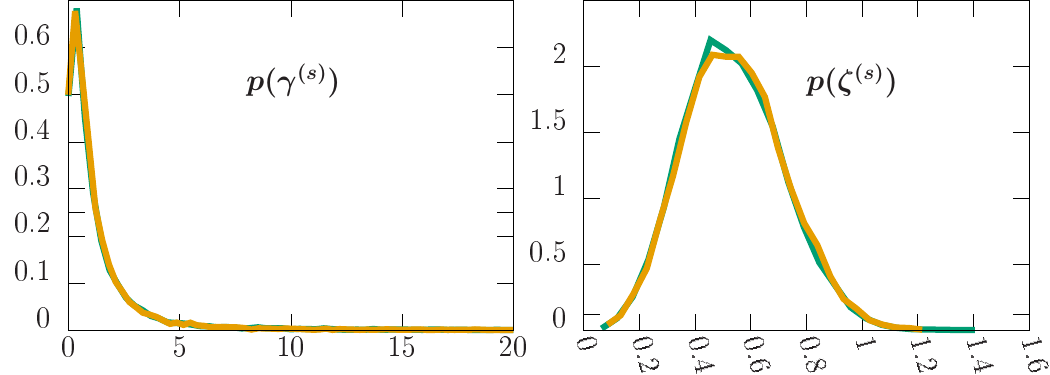}
    \caption{Left panel: Probability density functions of the shape parameter ($\gamma^{(s)}$) for hot spots (green) and cold spots (orange) from $10^4$ full sky pure CMB maps at HEALPix resolution of $n_{side}=16$. The horizontal axis has been clipped at $\gamma^{(s)}=10$, else it ranges up to an order of $10^2$. Both curves show qualitatively similar behaviour. Right panel: Probability density functions of the strength parameter ($\zeta^{(s)}$) for hot spots and cold spots. Both $p(\zeta^{(h)})$ and $p(\zeta^{(c)})$ behave similarly.}
    \label{estim_dist}
\end{figure}

Further, to estimate the distributions of values of these estimators, we obtain $10^4$ full sky statistically isotropic Gaussian random realisations of the pure CMB using the $\Lambda CDM$ concordance model based on the Planck 2018 best-fit theoretical angular power spectrum \citep{pl2018}. All pure CMB maps are at a HEALPix \citep{2005ApJ...622..759G} resolution of $n_{side}=16$ with $\ell_{max}=32$. We identify the hot and cold spots of the pure CMB maps with the help of the HEALPix F90 facility called `hotspot'. The probability densities of isotropy estimators for these full sky pure CMB maps are shown in Figure \ref{estim_dist}.
From the left panel of Figure \ref{estim_dist}, we see that pure CMB realisations exhibit a wide range of values of the shape parameter ($\gamma^{(s)}$) and the probability densities of $\gamma^{(s)}$ for both hot and cold spot placements behave similarly. From the right panel of Figure \ref{estim_dist}, we see that probability density functions of strengths of non-uniformity ($\zeta^{(s)}$) of both hot and cold spots are also similar in behaviour as expected.

\section{Analysis of toy models}\label{toy}

We test our estimators with two toy CMB maps which have been constructed to illustrate girdling and clustering of spots. Each of the toy maps have different strengths of non-uniformity. We use HEALPix for constructing and analysing these maps. The maps are at a resolution of $n_{side}=16$. 

\begin{figure}[tbp]
    \centering
    \includegraphics[width=\textwidth]{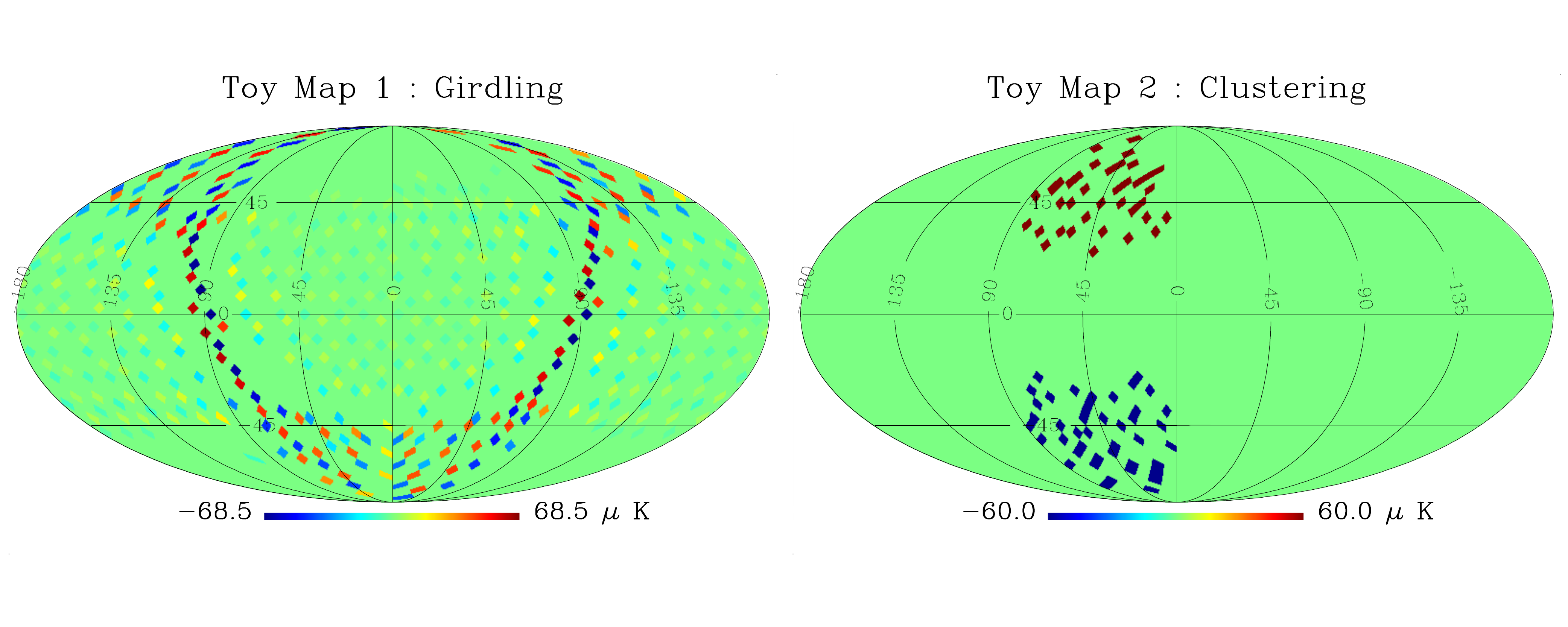}
    \caption{Two toy model maps of hot spots and cold spots, to show their girdled and clustered distributions, respectively, with different extents of non-uniformity or anisotropy.}
    \label{toymaps}
\end{figure}

We form the first toy map (TM1)  by constraining its spherical harmonic coefficients ($a_{\ell m}$) in the harmonic space. For this purpose, we randomly set the $a_{\ell m}$ for $\ell = 8$ to a value of $6$ and for $\ell = 28$ to to a value of $-6$ for both real and imaginary parts. Rest of the spherical harmonic coefficients are set to zero.  We then convert these $a_{\ell m}$ coefficients to HEALPix map at $n_{side} = 16$.


We construct the second toy map (TM2) in the following manner. To assign clustered hot and cold spots, we specify some positive and negative values for two groups of neighbouring pixels of a map array (say $m_a$) while setting other pixels to a value of zero. We also consider a randomly generated map $m_b$. Then TM2 is a linear combination of these maps, given by $m_a+\lambda\times m_b$. Here $\lambda$ is a very small fraction which highly suppresses spots from the randomly generated part ($m_b$), making them almost invisible.

We show the maps TM1 and TM2 with their hot spots and cold spots in a standard Mollweide projection in Figure \ref{toymaps}. From the left panel of Figure \ref{toymaps} we see that TM1 contains an approximately girdled distribution of spots. Since the non-uniformity of spots in TM1 is distinctly visible, the strength parameter should definitely be greater than zero but it may not be very high due to the presence of several weak (lightly coloured) hot and cold spots in the map. In the right panel of Figure \ref{toymaps} we see that the hot and cold spots of TM2 in the northern and southern hemispheres are quite clustered, and their non-uniformity is stronger relative to TM1.


\begin{table}
    \centering
    \caption{Values of the isotropy estimators for toy model maps. For the first toy map (TM1), $\gamma^{(s)}<1$ indicates girdled hot and cold spots, with strengths of $\zeta^{(s)}>1$. The second toy map (TM2) is more non-uniform with $\zeta^{(s)}>2$, and has clusters as $\gamma^{(s)}>1$.}
    \label{table1}
    \vspace{0.4cm}
\begin{tabular}{|c|c|c|c|} 
\hline 
Toy Map & Spots(s) & $\gamma^{(s)}$ & $\zeta^{(s)}$ \\ [1ex] 
 \hline
1 & h & $0.12690120$ & $1.2989503$ \\ [1ex]
1 & c & $0.12694086$ & $1.1855485$  \\ [1ex]
2 & h & $7.1326062$ & $2.7095464$  \\ [1ex]
2 & c & $8.4651048$ & $2.5985995$ \\ [1ex] 
 \hline
\end{tabular}
\end{table}

The values of the estimators ascertained from these two toy maps are given in Table \ref{table1}. Clearly, the strength of non-uniformity for TM1 is lower than that of TM2. The values of $\gamma^{(s)}$ for TM1 are lesser than unity, indicating girdles, and those for TM2 indicate clusters as those are greater than unity.

Both maps TM1 and TM2 contain several faint hot and cold spots which have low peak values. Despite being faint, some of these spots are visible for TM1 but for TM2 they are highly suppressed. These spots are numerous and their overall distributions are reasonably free from any signal of girdling or clustering. If such faint spots are considered on an equal footing with the dark coloured spots which have higher peak values, our estimators may not be able to correctly recover the signal of non-uniformity. This would be the case when all spots are treated as unit masses. On such a treatment, we find that
\begin{enumerate}
    \item[(a)]shape parameters of TM1 are $\gamma^{(h)}=1.2926231$, $\gamma^{(c)}=1.5587367$, neither of which correspond to girdles.
    \item[(b)] For TM2, we have $\gamma^{(h)}=0.98292051$, $\gamma^{(c)}=0.47354657$, neither of which correspond to clusters.
    \item[(c)] Strength parameters of TM1 are $\zeta^{(h)}=0.34948347$, $\zeta^{(c)}=0.38412180$, which indicate very weak non-uniformity.
    \item[(d)] For TM2, the strengths are $\zeta^{(h)}=0.19635766$, $\zeta^{(c)}=0.17391740$, which indicate nearly uniform placement of spots.
\end{enumerate}
These estimator values obtained with unit-mass weights are markedly insensitive to the visible signatures of non-uniformity of spots. Hence, the use of non-unit mass weights in the orientation matrix for CMB spots is indispensable for a reliable recovery of signals of non-uniformity.


\section{Application on Foreground Minimized CMB maps}\label{analysis_data}

We simulate $10^4$ pure CMB maps using the Planck 2018  best-fit theoretical angular power spectrum \citep{pl2018} at $n_{side}= 16$  with $\ell_{max} = 32$. All these maps contain pixel smoothing corresponding to the  pixel resolution $n_{side} = 16$. The advantage of pixel window smoothing is two fold. First,  if the maps are not smoothed by the window function unwanted errors in the characterization of peaks may occur. Secondly, since the HEALPix pixel tessellation  does not follow an isotropic distribution, systematics can be introduced in the shape and strength parameters if the maps are not smoothed by the pixel window functions. For the observed CMB maps we downgrade them from $n_{side} = 2048$ (for Planck) or $n_{side}=512$ (for WMAP) to $n_{side}= 16$ using ‘ud\_grade’ facility. Therefore, these low resolution data maps also correctly contain pixel window smoothing effects. We convert the pixel smoothed data maps to spherical harmonic coefficients and reconstruct the actual data maps for our analysis by taking multipoles between $\ell = 2$ to $32$. In addition, any existing beam smoothing effects are removed from observed CMB maps. Thus similar to pure CMB maps, the observed maps are not convolved with any beam window function.

We exclude multipoles $\ell=0,1$ from our analysis since these correspond respectively, to the monopole of uniform CMB temperature \citep{Fixsen_2009} and the dipole due to our motion relative to the CMB rest frame \citep{Bucher:2015eia}. We neglect noise in the analysis, as it is expected to be insignificant for the low multipole range maps analysed here \citep{Tegmark_2000}, as these correspond to large angular scales. We utilise the F90 facility `hotspot' of HEALPix for finding the hot and cold spots for simulated as well as observed CMB maps.

We compute the values of the shape and strength parameters, denoted by $x$ for each of the observed and simulated maps. Then the fraction $P^t(x)$ for each observed data map can be calculated by counting the number of simulations which have a value of $x_{sim}$ greater than that from observed data ($x_{data}$) and dividing the same by the total number of simulations. Conventionally, $x_{data}$ for which $P^t(x)$ are found outside the confidence interval bounded by the probability values $0.05$ and $0.95$ are considered unlikely relative to pure CMB realisations. Thus for $P^t(x)<5\%$, this would imply that $x_{data}$ is unusually high, whereas $x_{data}$ becomes unusually low for $P^t(x)>95\%$.

In the following Sections \ref{sans_gal} and \ref{partial}, we describe the observed CMB maps and masks used for the analysis. We first choose to include the galactic region in the analysis along with the other parts of the sky. In this case, our study concerns two cases. First we use the entire sky and secondly we exclude only the region corresponding to the non-Gaussian cold-spot (NGCS). Thereafter, we exclude the galactic regions from the analysis. The inclusion and exclusion of the galactic regions in two different analyses help us understand effects of any (minor) residual foregrounds that may be present in the galactic regions even after performing foreground minimization. We present the results for isotropy statistics in Section \ref{results}.

\subsection{Case I: Without galactic masks}\label{sans_gal}

\subsubsection{Input maps}\label{fullSky}
\begin{figure}[tbp]
    \centering
    \includegraphics[width=1.0\textwidth]{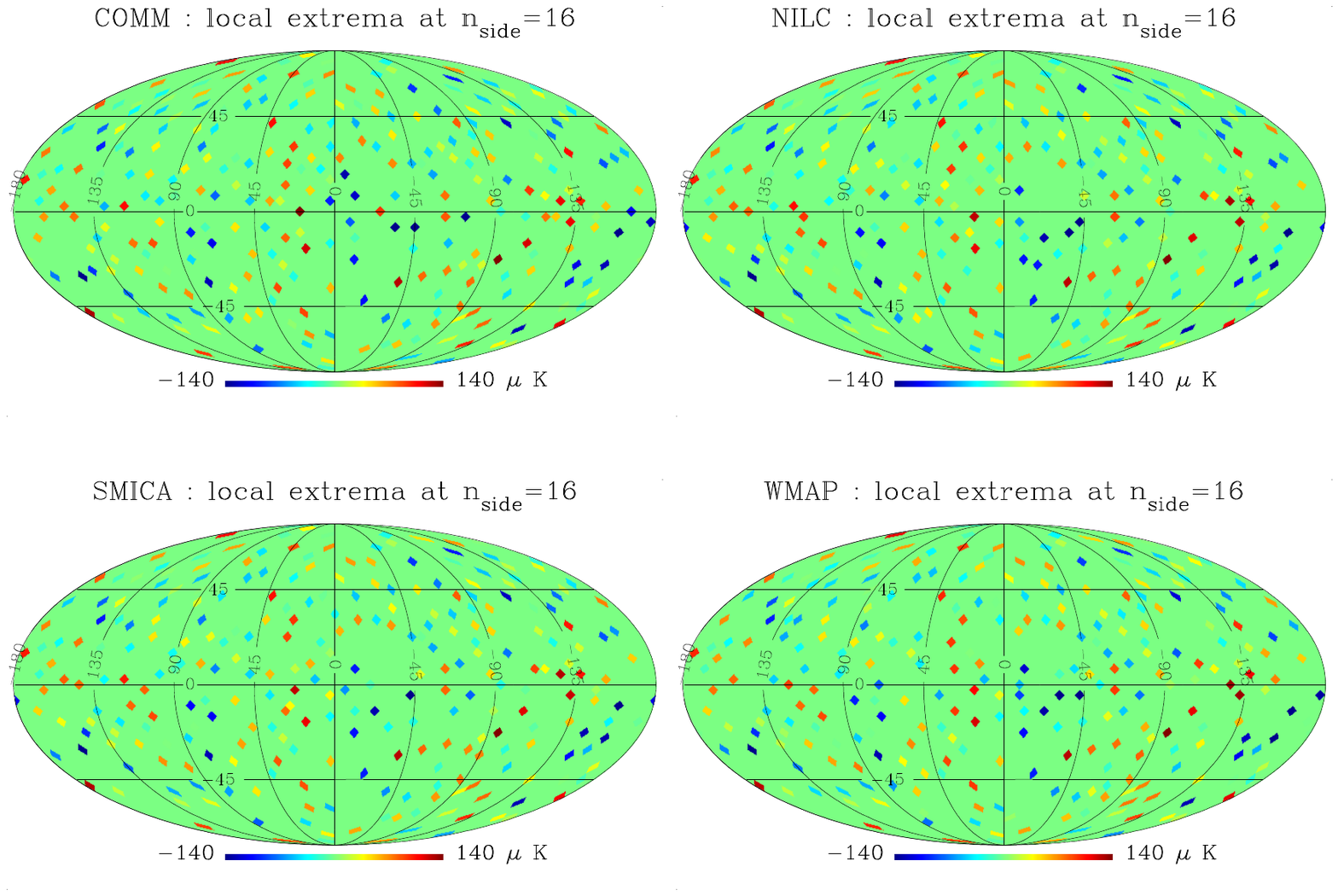}
    \caption{Mollweide projections of hot spots and cold spots for the four full sky cleaned CMB maps, i.e., COMM, NILC, SMICA, and WMAP at a HEALPix resolution of $n_{side}=16$. These subfigures illustrate how hot and cold spots are placed on observed full sky CMB maps.}
    \label{extrema_maps}
\end{figure}

We use four cleaned maps, namely, those of the 2018 release \citep{akrami_planck_2020-1} of Planck's Commander (COMM), NILC, and SMICA, and WMAP's 9 year ILC \citep{Bennett_2013} (hereafter referred to as WMAP) for the full sky analysis. We present Mollweide projections showing hot and cold spots of these four individual full sky maps in Figure \ref{extrema_maps} to illustrate how the spots are scattered on the observed full sky CMB.

\subsubsection{NGCS mask}\label{maskNGCS}

\begin{figure}[tbp]
    \centering
    \includegraphics[width=0.5\textwidth]{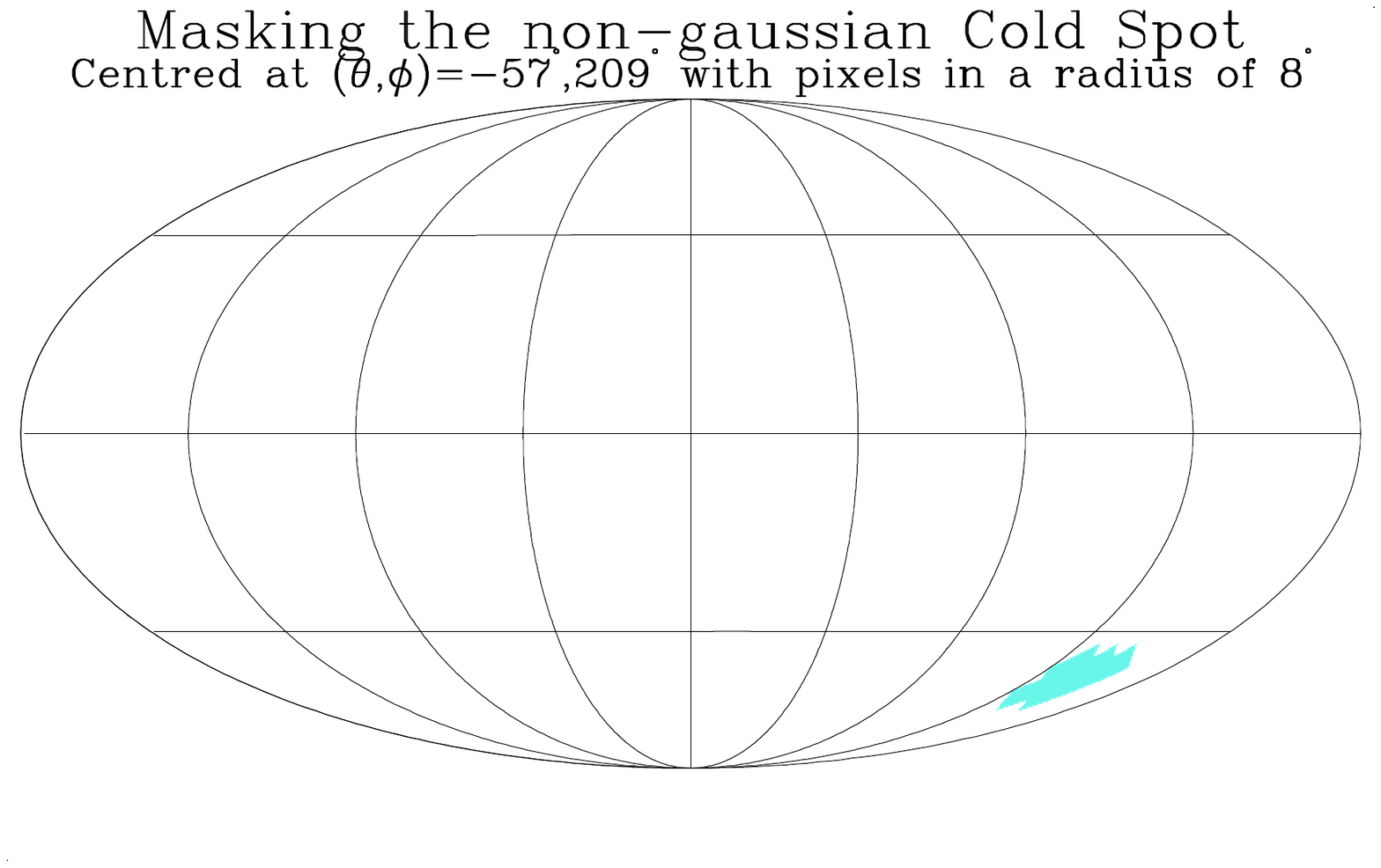}
    \caption{The non-Gaussian cold spot (NGCS) mask at a HEALPix resolution of $n_{side}=16$: the masked region is in cyan; the white region is unmasked.}
    \label{cs}
\end{figure}

The non-Gaussian Cold spot (NGCS) \citep{10.1111/j.1365-2966.2006.10312.x, MARTINEZGONZALEZ2006875, vielva_2010} is a well established anomaly, centred at $(\theta,\phi)=\ang{-57},\ang{209}$. The north-south power asymmetry \citep{Bernui_2014, Quartin_2015, Akrami_2014, Eriksen_2004, Eriksen_2004b, Eriksen_2007} was seen to be correlated with the NGCS \citep{PhysRevD.80.123010}, the significance of which effect was seen to be low for low resolution \citep{10.1111/j.1365-2966.2010.16905.x} maps at HEALPix $n_{side}=16$. In order to check for any correlation between the isotropy estimators and the NGCS, we mask out the map pixels in a radius of $\ang{8}$ from the cold spot center. This mask is shown in Figure \ref{cs}. 
This mask is used independently for the four cleaned maps and all pure CMB maps described above, and later along with two galactic masks for nine observed maps as mentioned below.

\subsection{Case II: With galactic masks}\label{partial}

\subsubsection{Input maps}\label{galmaps}

In addition to the four cleaned CMB maps (COMM, NILC, SMICA, and WMAP 9 year ILC), we use WMAP's foreground reduced Q, V and W frequency maps and Planck's `fgsub-sevem' $\SI{70}{\giga \hertz}$ and $\SI{100}{\giga\hertz}$ frequency maps. These five frequency maps will hereafter be referred to as freqQ, freqV, freqW, freq70, and freq100, respectively.

\subsubsection{Masks}\label{galmsk}
Foreground residuals predominantly in the galactic region of cleaned CMB maps may cause certain unusual patterns to manifest in the observed data when it is analysed relative to pure CMB realisations. The use of a mask for the galactic region and some extra-galactic point sources is required to check if such unusual patterns are truly characteristic of the CMB or due to residual systematics.

\begin{figure}[htbp]
    \centering
    \includegraphics[width=\textwidth]{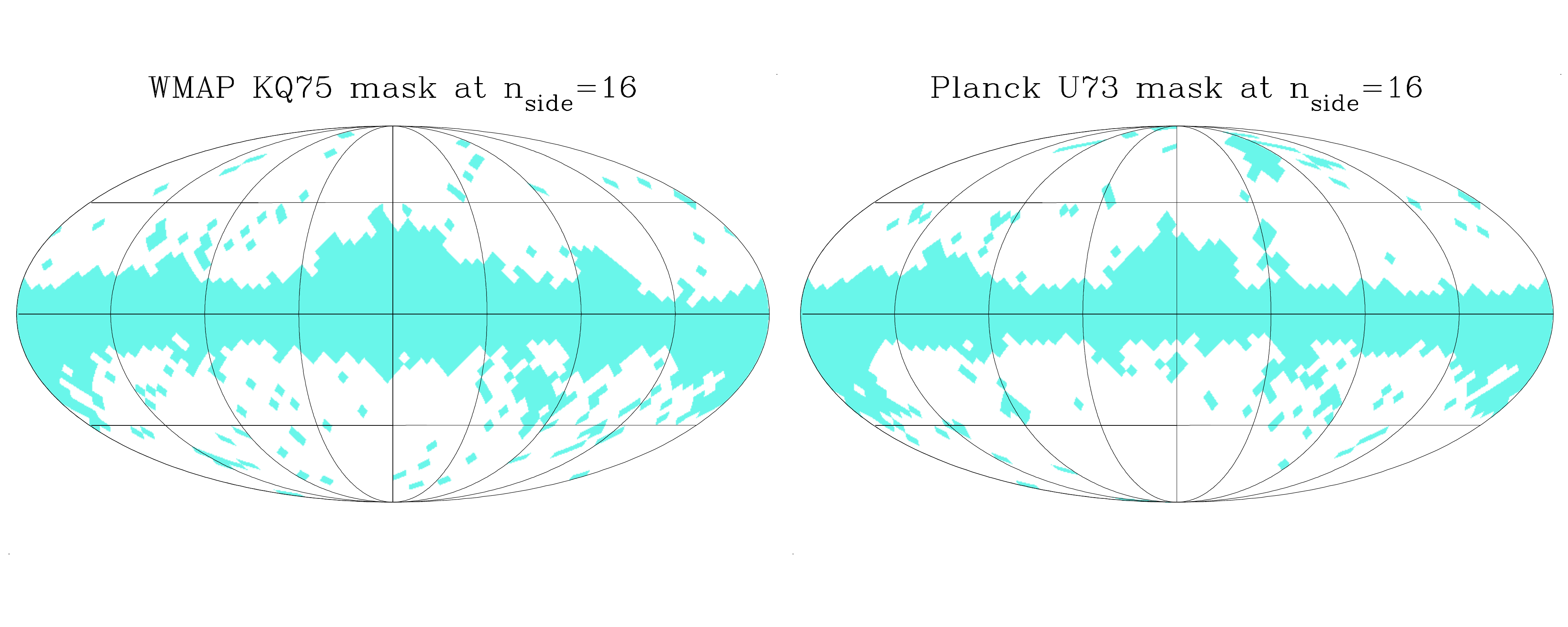}
    \caption{$KQ75$ and $U73$ masks at a HEALPix resolution of $n_{side}=16$: the masked region is in cyan; the white region is unmasked.}
    \label{masks}
\end{figure}

We utilise low resolution versions of the $KQ75$ mask of WMAP 9 year data and the $U73$ mask of Planck 2018 data, which is a product of the temperature confidence masks associated with COMM, NILC, SEVEM, and SMICA. The two masks are shown in Figure \ref{masks}. The $KQ75$ mask is very conservative in the sense that it comprises a wider galactic cut and conceals a larger number of point sources as compared to the $U73$ mask.

These low resolution masks are obtained by downgrading their high resolution variants to HEALPix $n_{side}=16$ and applying thresholds of $0.85$ and $0.98$ for the $KQ75$ and $U73$ masks, respectively. A threshold of $y$ entails setting all pixels with values $\leq y$ to $0$, and the rest to $1$, after a mask is downgraded. The resulting sky fractions for the $KQ75$ and $U73$ masks are $62.9\%$ and $67.5\%$, respectively.  The thresholds chosen here are a bargain between a good sky fraction of the CMB for signal detection vis a vis any dominant foreground sources at the large scales \citep{Tegmark_2000} considered here. The choices for thresholds are not altogether arbitrary, but inspired from \cite{_iso_stat}. Additionally, we consider the union masks of $KQ75$ and $U73$ with the NGCS mask, referred to as $KQ75-CS$ and $U73-CS$, respectively. 

We apply these galactic masks one by one, simultaneously to the four cleaned CMB and five foreground reduced frequency maps and $10^4$ pure CMB maps. The partial sky analysis with the nine input maps of observed CMB using galactic masks provides checks of robustness of any detected signal, against the following:
\begin{enumerate}
    \item Different galactic masks ($KQ75$ and $U73$),
    \item Various methods of foreground cleaning (Planck's COMM, NILC, and SMICA, and WMAP's ILC) and reduction (foreground template model reduction \citep{Hinshaw_2007_cleaning_algorithm, Bennett_2013} for freqQ, freqV, freqW and SEVEM \citep{10.1111/j.1365-2966.2011.20182.x, leach_component_2008} for fgsub-sevem freq70 and freq100 maps),
    \item Several frequencies (Q, V, W bands and $\SI{70}{\giga\hertz}, \SI{100}{\giga\hertz}$),
    \item Presence or absence of the NGCS when the galactic region is masked out, and
    \item Different instruments (WMAP and Planck's Low Frequency \citep{_lfi} and High Frequency Instruments \citep{aghanim_planck_2020}).
\end{enumerate}

\section{Results and Analysis}\label{results}
We present results from the analysis of foreground minimized CMB maps with and without galactic masks. Without galactic masks, we analyse four cleaned CMB maps, firstly, for the full sky, and secondly after masking the NGCS. With galactic masks, we analyse nine observed CMB maps.

\subsection{Case I: Without galactic masks}

 For the four full sky cleaned CMB maps used, we tabulate the values of the estimators in Table \ref{table2}. We see that mostly uniform placements of hot spots and and cold spots can be inferred from the values of the strength parameter. 
 \begin{table}
    \centering
    \caption{Values of the isotropy estimators for observed full sky CMB maps: Strengths $\zeta^{(h)}$ and $\zeta^{(c)}$ are low, indicating mostly uniform placements of hot spots and cold spots.}
    \label{table2}
    \vspace{0.4cm}
\begin{tabular}{|c|c|c|c|c|} 
\hline 
Data Map & $\gamma^{(h)}$ & $\zeta^{(h)}$ & $\gamma^{(c)}$ & $\zeta^{(c)}$  \\ [1ex] 
 \hline

       COMM &     $1.5660$  &    $0.2137$ &     $1.5079$ &     $0.3479$  \\[1ex]
       NILC &    $1.5811$ &     $0.1849$  &    $0.7978$ &      $0.2800$  \\[1ex]
       SMICA &    $1.0435$ &     $0.2089$ & $0.5460$ &     $0.2367$  \\[1ex]
       WMAP &   $0.4239$ &     $0.2148$   &  $2.4463$ &    $0.3408$   \\[1ex]

 \hline
\end{tabular}
\end{table}

\begin{figure}[tbp]\centering
    \includegraphics[width=\textwidth]{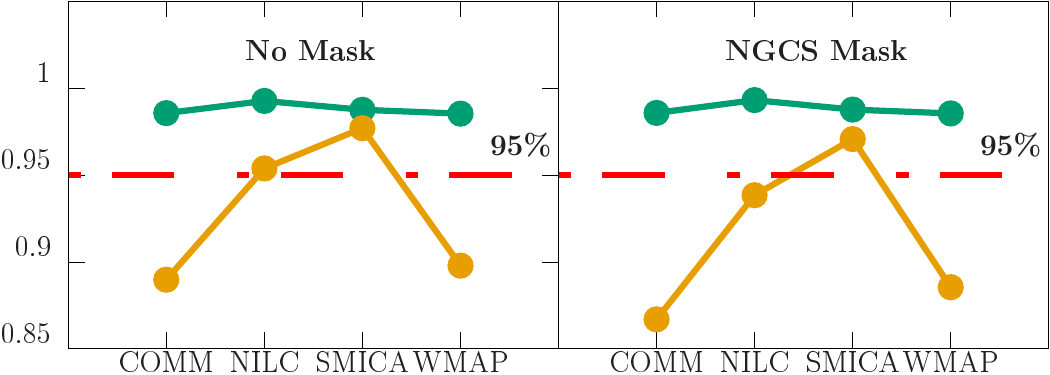}
    \caption{Left panel: $P^t(\zeta^{(h)})$ (green) and $P^t(\zeta^{(c)})$ (orange) are shown; for the four full sky maps, unusual uniformity of hot spots is seen for all maps, and that of cold spots is seen for NILC and SMICA. Right panel: for the four maps with the NGCS mask, unusual uniformity of hot spots persists for all maps, but that of cold spots is seen for SMICA.}
    \label{pval1}
\end{figure}

In Figure \ref{pval1}, we show $P^t$ for the four cleaned CMB maps, without galactic masks. On the left panel of Figure \ref{pval1}, results from the full sky maps are presented. We find that the strength parameters $\zeta^{(h)}$ and $\zeta^{(c)}$ are relatively lower than those from pure CMB simulations, as $P^t(\zeta^{(h)})$ ranges between $98.52\%$\textendash$99.25\%$ and $P^t(\zeta^{(c)})$ ranges between $88.97\%$\textendash$97.68\%$. Thus, we see robustly low $\zeta^{(h)}$ for all the four cleaning methods. The lowest value of $\zeta^{(h)}=0.1849$ with $P^t(\zeta^{(h)})=99.25\%$ is seen for NILC. Low $\zeta^{(c)}$ for NILC and SMICA are seen, of which the lowest $\zeta^{(c)}=0.2367$ with $P^t(\zeta^{(c)})=97.68\%$ is seen for SMICA.

On the right panel of Figure \ref{pval1}, results from the four maps after masking the NGCS are shown.  We find that $P^t(\zeta^{(h)})$ ranges between $98.53\%$\textendash$99.30\%$, and $P^t(\zeta^{(c)})$ lies between $86.69\%$\textendash$97.06\%$. Thus removal of the NGCS very slightly increases the significance of low $\zeta^{(h)}$ while decreasing the significance of low $\zeta^{(c)}$ relative to pure CMB realisations. Robustly low $\zeta^{(h)}$ for all the four cleaned maps is seen with the NGCS mask, and the lowest value of $\zeta^{(h)}=0.1849$ with $P^t(\zeta^{(h)})=99.30\%$ occurs for NILC. Low $\zeta^{(c)}$ is seen for SMICA with $\zeta^{(c)}=0.2516$ and $P^t(\zeta^{(c)})=97.06\%$.

Thus, unusually weak non-uniformity of hot spots in all four maps is seen for full sky as well as for partial sky outside the NGCS mask. However, the shape parameters $\gamma^{(h)}, \gamma^{(c)}$ are in good agreement with pure CMB realisations. Unusually low strength of non-uniformity of cold spots is seen for NILC and SMICA for full sky. Interestingly, after masking the galactic regions, this signal spreads over several cleaned maps. We discuss this in detail in the subsequent section. 

\subsection{Case II: With galactic masks}\label{robust}

\begin{figure}[htbp]
    
    \includegraphics[width=\textwidth]{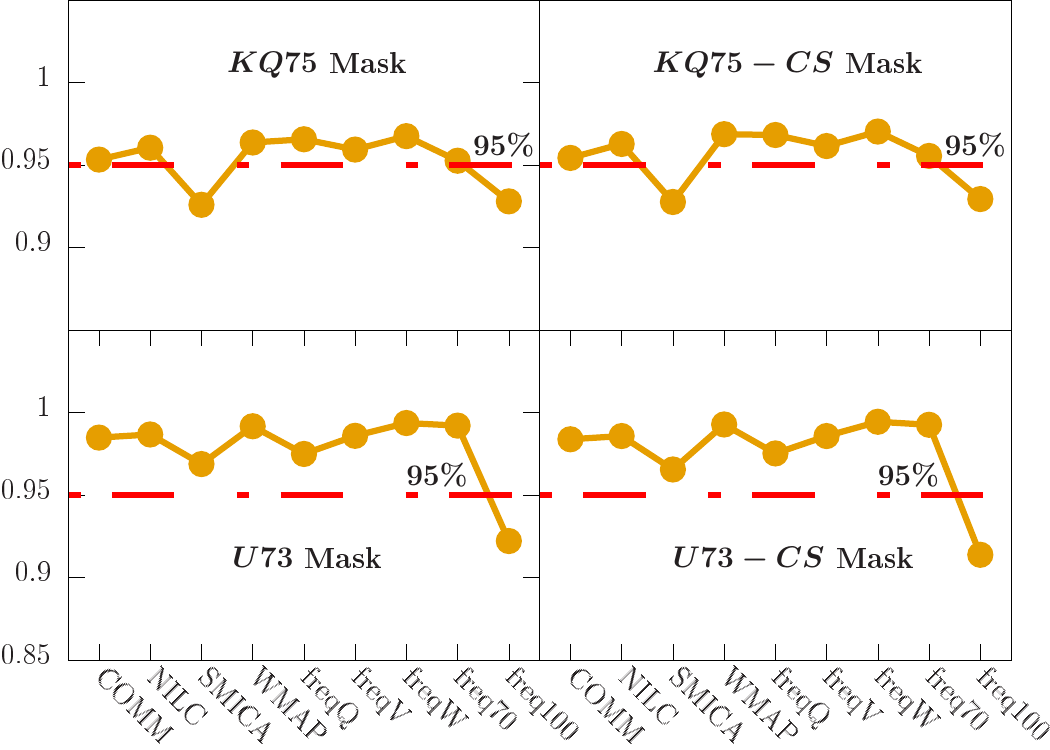}
    \caption{Top left panel: $P^t(\zeta^{(c)})$ is in orange; $\zeta^{(c)}$ is significantly low for all maps except SMICA and freq100. Top right panel: unusual uniformity of cold spots is seen for all maps except SMICA and freq100. Bottom left panel: similarly unusual uniformity of cold spots is seen for all maps except freq100. Bottom right panel: again, $\zeta^{(c)}$ is significantly low for all maps except freq100.}
    \label{p1combo}
\end{figure}

In Figure \ref{p1combo}, we show $P^t$ for the nine observed partial sky CMB maps, with galactic masks. On the top left panel of Figure \ref{p1combo}, results obtained with the use of $KQ75$ mask are shown. Significantly low $\zeta^{(c)}$ is seen for all the maps, except SMICA and freq100. Lowest $\zeta^{(c)}=0.4007$ with $P^t=96.75\%$ is seen for freqW, without the NGCS mask. On the top right panel of Figure \ref{p1combo}, we again obtain robust results of anomalously low $\zeta^{(c)}$ for the $KQ75-CS$ mask for all maps, except SMICA and freq100. Lowest $\zeta^{(c)}=0.3924$ with $P^t=97.03\%$ is seen for freqW.
 
On the bottom left panel of Figure \ref{p1combo}, results obtained with the use of $U73$ mask  are shown. Again, an unusually low $\zeta^{(c)}$ is seen robustly for various maps. Significantly low $\zeta^{(c)}$ is seen for all maps except freq100. Lowest value of $\zeta^{(c)}=0.2446$ occurs for freqW with $P^t(\zeta^{(c)})=99.37\%$ without the NGCS mask. From the bottom right panel of Figure \ref{p1combo}, for $U73-CS$ mask, we see significantly low $\zeta^{(c)}$ for all maps except freq100. Lowest value of $\zeta^{(c)}=0.2432$ with $P^t=99.44\%$ occurs for freqW.

We have presented the numerical values of $P^t(\zeta^{(c)})$ in Table \ref{tcombo}. These numerical values indicate that the unusually weak non-uniformity of cold spots is more significant with the use of the less conservative $U73$ and $U73-CS$ masks. The values of $P^t$ in Table \ref{tcombo} reveal that the the absence of the NGCS slightly complements the unusual nature of $\zeta^{(c)}$ with the $KQ75$ mask.

\begin{table}
 \centering   
    \caption{Numerical values of $P^t(\zeta^{(c)})$ for partial sky analysis with galactic masks. A very high $P^t$ entails that the value of the estimator from observed data is unusually low, and vice versa. Hence, we see anomalously low $\zeta^{(c)}$ for several maps, albeit with different $P^t$ values for different masks. Significant values of $P^t$ are in boldface.}
    \label{tcombo}
    \vspace{0.4cm}
\begin{tabular}{|c|c|c|c|c|}
\hline
   Map  & $KQ75$  & $KQ75-CS$  & $U73$  & $U73-CS$ \\ [1ex] 
 \hline
      COMM     &       $\mathbf{0.9533}$  &     $\mathbf{0.9543}$ &      $\mathbf{0.9848}$ &      $\mathbf{0.9838}$ \\[1ex]
       NILC     &       $\mathbf{0.9605}$  &     $\mathbf{0.9628}$ &      $\mathbf{0.9867}$ &      $\mathbf{0.9858}$ \\[1ex]
       SMICA    &       $0.9259$  &     $0.9276$ &      $\mathbf{0.9688}$ &      $\mathbf{0.9655}$ \\[1ex]
       WMAP     &       $\mathbf{0.9637}$  &     $\mathbf{0.9687}$ &      $\mathbf{0.9917}$ &      $\mathbf{0.9928}$ \\[1ex]
   freqQ    &       $\mathbf{0.9656}$  &     $\mathbf{0.9682}$ &      $\mathbf{0.9749}$ &      $\mathbf{0.9752}$ \\[1ex]
      freqV    &       $\mathbf{0.9594}$  &     $\mathbf{0.9615}$ &      $\mathbf{0.9859}$ &      $\mathbf{0.9858}$ \\[1ex]
      freqW    &       $\mathbf{0.9675}$  &     $\mathbf{0.9703}$ &      $\mathbf{0.9937}$ &      $\mathbf{0.9944}$ \\[1ex]
      freq70   &       $\mathbf{0.9526}$  &     $\mathbf{0.9555}$ &      $\mathbf{0.9921}$ &      $\mathbf{0.9926}$ \\[1ex]
      freq100  &       $0.9280$  &     $0.9294$ &      $0.9221$ &      $0.9138$ \\[2ex]

\hline
\end{tabular}
\end{table}


Thus, unusually low strength of non-uniformity of cold spots is seen for all maps, except SMICA and freq100 when $KQ75$ mask is applied with and without the NGCS mask. Again, such weak non-uniformity of cold spots is seen for all maps, but with the exception of freq100, when $U73$ mask is applied with and without the NGCS mask. Strength of non-uniformity of hot spots ($\zeta^{(h)}$) is also low, but the effect is not significant relative to pure CMB realisations. A single instance of an unusual ring structure of hot spots for freqQ (not shown in Figure \ref{p1combo}) is seen with $P^t(\gamma^{(h)})=99.32\%$ for the $U73$ mask and $P^t(\gamma^{(h)})=99.33\%$ for the $U73-CS$ mask. But no significantly unusual $\gamma^{(h)}$ was seen with the more conservative $KQ75$ and $KQ75-CS$ masks. This implies that the ring structure of hot spots for freqQ could be due to some foreground residuals when the $U73$ and $U73-CS$ masks are applied. Apart from this occurrence, no other unusual clustering or girdling is seen for any of the maps with the four masks.

\section{Is anomalously weak non-uniformity due to low variance?}\label{quadoct}

It has been seen before that on an average hot and cold spots of the observed CMB have unexpectedly low peak values \citep{Larson_2004}. Besides, the variance of the CMB temperature anisotropy field is anomalously low \citep{10.1111/j.1365-2966.2008.13149.x}. Therefore such low variance in addition to low mean values of local extrema entails that the peak values when measured will turn out to be lower than expected from statistically isotropic Gaussian random fluctuations of the temperature field. Since peak values of spots are incorporated as non-unit mass weights in the orientation matrix, our novel isotropy statistics carry this information. Further, the low variance anomaly was seen to be confined to the northern ecliptic hemisphere. Any such directional preference is additionally manifested in the relative magnitudes of eigenvalues, which is encapsulated by the strength of non-uniformity ($\zeta^{(s)}$). Thus it is important to check if the signal of anomalously low strength of non-uniformity as seen for hot spots on full sky and partial sky with the NGCS mask, or that for cold spots on partial sky outside galactic masks are correlated with the low CMB variance anomaly. Another interesting question to consider is that of how the shape and strength parameters behave on different scales. We can understand this behaviour by analysing CMB maps containing different ranges of multipoles.

Thus, we seek to investigate the following: 
\begin{enumerate}
\item[(a)] whether the signal of anomalously weak non-uniformity is correlated with low CMB temperature variance, and
\item[(b)] how the isotropy statistics behave on different scales.
\end{enumerate}
Since the CMB low variance anomaly disappears when the quadrupole and octupole are excluded \citep{10.1111/j.1365-2966.2010.18067.x}, therefore both these questions can be addressed by excluding the quadrupole and octupole which correspond to two of the largest scales.

\begin{figure}
    \centering
    \includegraphics[width=0.5\textwidth]{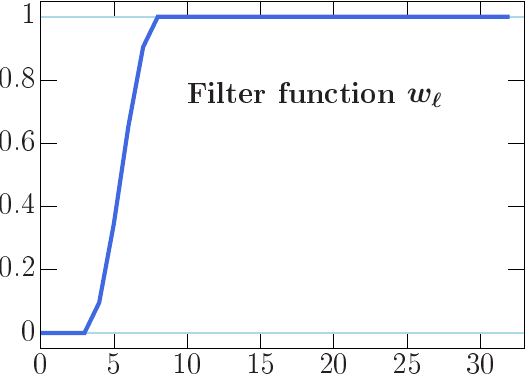}
    \caption{The cosine filter function which is multiplied with the spherical harmonic coefficients of CMB maps to study how quadrupole-octupole contributions and hence low CMB variance may affect the strength of non-uniformity of hot and cold spots.}
    \label{fltr}
\end{figure}

Thus we reconstruct all pure and observed CMB maps after multiplying their spherical harmonic coefficients by a cosine filter,
\begin{eqnarray}
w_\ell&=&0 \quad \text{for $\ell\in [0,3]$}, \nonumber \\
w_\ell&=&\frac{1}{2}\left[1-\cos{\left(\frac{\pi(\ell-3)}{5}\right)}\right] \quad \text{for $\ell\in[4,7]$}\nonumber, \\
w_\ell&=&1 \quad\text{for $\ell\in[8,32]$}.
\end{eqnarray}
This filter function (shown in Figure \ref{fltr}) excludes the quadrupole and octupole and smoothly suppresses low multipoles upto $\ell=7$.
We perform the same analysis as in Section \ref{results} for the new maps and find that:
\begin{enumerate}
    \item[1.] without galactic masks, the signal robustly disappears across all cleaned maps for the full sky and with NGCS mask. This is reflected in the values of $P^t$, which range between $48.68\%$\textendash $90.92\%$ for $\zeta^{(h)}$ and $30.12\%$\textendash $94.01\%$ for $\zeta^{(c)}$.
    \item[2.] However, with galactic masks, the signal does not persist in any of the foreground minimized maps except maps of freqQ (with $U73$ mask) and freqV (with $KQ75, KQ75-CS$ masks) for which $\zeta^{(c)}$ is unusually low. Thus we see that $P^t(\zeta^{(h)})$ ranges between $31.22\%$\textendash $81.59\%$ whereas $P^t(\zeta^{(c)})$ lies between $47.93\%$\textendash $97.02\%$.
\end{enumerate}
Since the signal of weak non-uniformity robustly disappears across all cleaned maps for full and partial sky coverage, this indicates that the signal could potentially be related with low CMB variance and quadrupole-octupole contributions, and these may share a common origin. Besides, the loss of robustness in disappearance of the signal occurs only for freqQ and freqV maps outside the galactic region. This strongly suggests that some foreground residuals in these two maps contribute to the signal.

\section{Some other considerations}\label{h+c}

In our work we have considered separate sets of hot spots and cold spots. Further we have taken into account the peak values of spots relative to the mean CMB temperature, in the form of non-unit mass weights. In Section \ref{toy}, we have demonstrated for two toy maps how the use of unit mass weights is insufficient to accurately recover any signal of non-uniformity of spots. However, for pure and observed CMB maps, one may be curious to consider:
\begin{enumerate}
    \item[(a)] The use of unit masses for the hot and cold spots,
    \item[(b)] The composite set of hot and cold spots taken together.
\end{enumerate}
We attempt at shedding light on the consequences of these choices by analysing both types of spots together, using a single orientation matrix, 
\begin{equation}
    \mathcal{T}^{(h+c)}=\frac{\sum_i m^{(h)}_i \tau^{(h)}_i+\sum_j m^{(c)}_j \tau^{(c)}_j}{\sum_i m^{(h)}_i+\sum_j m^{(c)}_j},
\end{equation} for two cases, i.e, with unit and non-unit masses. Here, the superscript $^{(h+c)}$ denotes the composite set including both hot and cold spots.

When both kinds of spots are taken together, we expect a more uniform (isotropic) placement of the spots from pure CMB realisations, compared to the case when they are considered separately. Thus we expect low strength of non-uniformity ($\zeta^{(h+c)}$) for most of the pure CMB maps. A low $\zeta^{(h+c)}$ corresponds to very closely spaced eigenvalues, so that there are almost no preferred directions for any non-uniform placements of the spots.

\begin{figure}[htbp]
\centering
    \includegraphics[width=\textwidth]{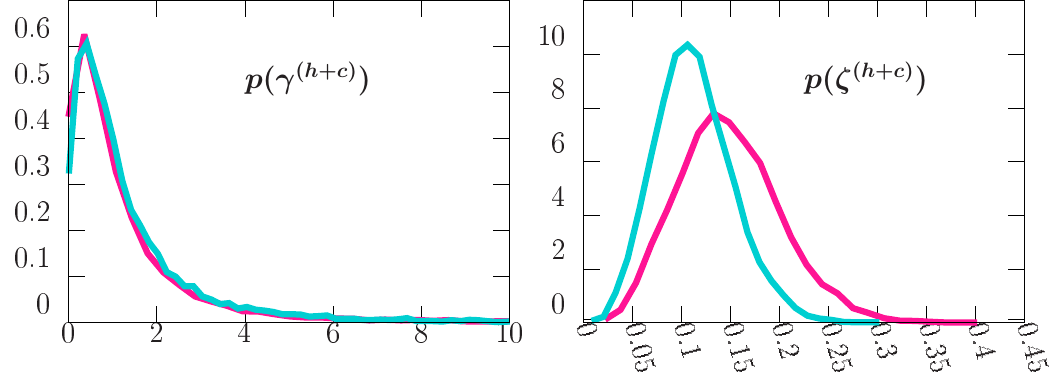}

    \caption{Left panel: Probability density functions  of the shape parameter ($\gamma^{(h+c)}$) obtained from Monte-Carlo simulations are shown for unit masses (magenta) and for non-unit masses (cyan). The horizontal axis is clipped at 10, else it ranges up to an order of $10^2$. Both the densities show qualitatively similar behaviour. Right panel: Probability density functions of the strength parameter ($\zeta^{(h+c)}$) for unit-mass and non-unit mass weights are shown. For non-unit mass case $p(\zeta^{(h+c)})$ peaks at lower values of $\zeta^{(h+c)}$.}
    \label{iso_mass}
\end{figure}

\begin{figure}[htbp]
\centering
    \includegraphics[width=\textwidth]{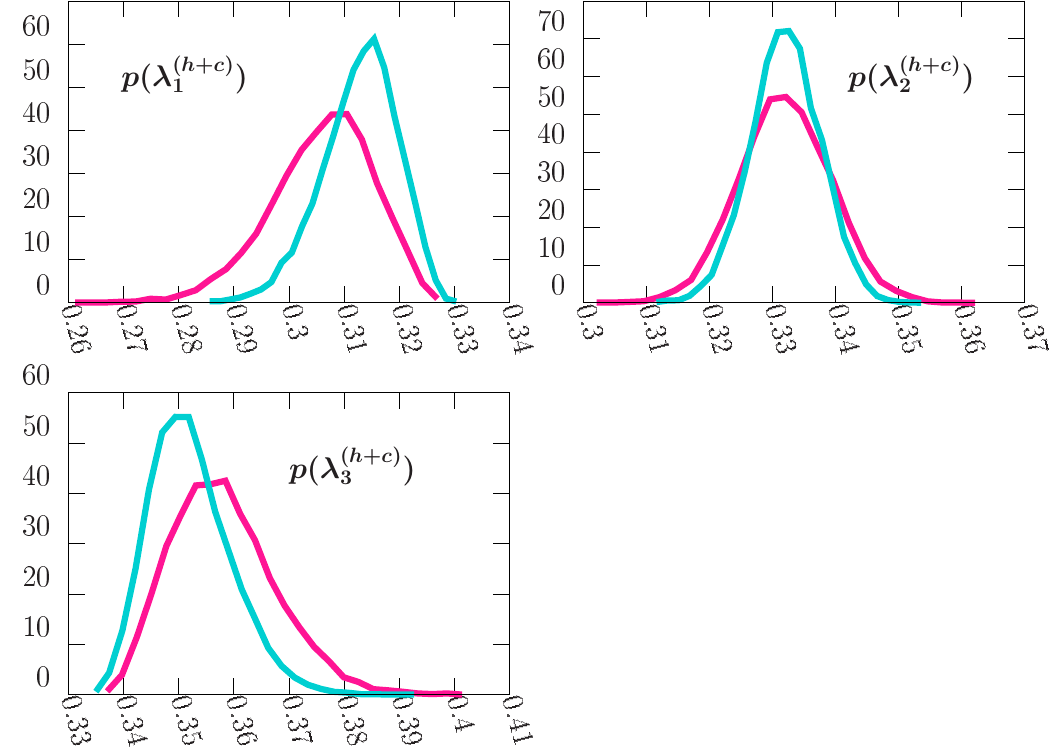}
    \caption{Top left panel shows probability density functions of the smallest eigenvalue ($\lambda^{(h+c)}_1$) in magenta for unit masses, and in cyan for non-unit masses. Top right panel shows the density functions of the intermediate eigenvalue ($\lambda^{(h+c)}_2$). Bottom panel shows the density functions of the largest eigenvalue ($\lambda^{(h+c)}_3$). All three density functions for non-unit mass weights have a smaller spread.  The most probable values of $\lambda^{(h+c)}_1$, $\lambda^{(h+c)}_2$, $\lambda^{(h+c)}_3$ for non-unit masses are closer to each other  and to $1/3$ as opposed to those for unit-masses. This is due to the additional randomness introduced into the orientation matrix while using non-unit mass weights.}
    \label{iso_mass2}
\end{figure}

We know that with the choice of non-unit masses as opposed to unit masses, we are introducing the randomness which is attributable to the peak values of spots in the system. This is in addition to the randomness associated with the eigenvector directions. Hence performing a comparative study of unit ($m_i^{(h)}, m_i^{(c)}=1$) and non-unit masses will help elucidate the advantage of considering non-unit masses to corroborate the expected isotropy of eigenvalues. We show the probability distributions of the shape and strength parameters for simulated pure CMB maps in Figure \ref{iso_mass} and those for the three eigenvalues ($\lambda_1^{(h+c)}\leq\lambda_2^{(h+c)}\leq\lambda_3^{(h+c)}$) in Figure \ref{iso_mass2}. We use the colour magenta for the case of unit-mass weights and cyan for the case of non-unit mass weights.

The left panel of Figure \ref{iso_mass} shows that the distributions for unit and non-unit masses behave similarly for the shape parameter $\gamma^{(h+c)}$. From the right panel of Figure \ref{iso_mass}, for $\zeta^{(h+c)}$ with non-unit masses, the range of values is more constricted relative to that with unit-masses, and the peak of the $p(\zeta^{(h+c)})$ curve is at lower values of $\zeta^{(h+c)}$ for non-unit mass weights, hence corresponding to greater uniformity. From subfigures of Figure \ref{iso_mass2}, we see that the expectations of uniformity are adhered to for unit and non-unit masses. However, in the case of non-unit masses, the most probable eigenvalues are definitely closer in magnitude to each other and to $1/3$. Further, the spread in the probability distributions of the eigenvalues is smaller for non-unit mass weights.

\begin{figure}[tbp]
    \centering
    \includegraphics[width=0.5\textwidth]{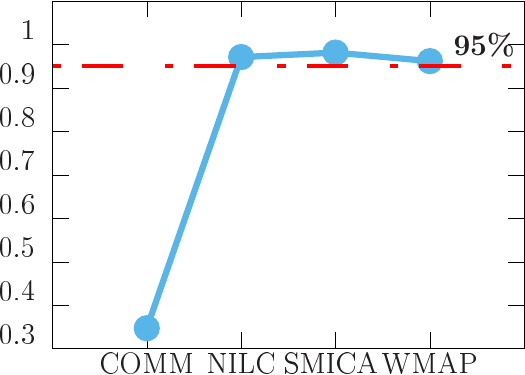}
    \caption{For non-unit mass weights, the fraction $P^t(\zeta^{(h+c)})$ is shown in light-blue. The observed full sky maps of NILC, SMICA and WMAP have unusually low values of $\zeta^{(h+c)}$.}
    \label{fig_h+c}
\end{figure}

As for observed CMB maps, we find that when both types of spots are taken together with unit masses, the observed data is in good agreement with the concordance model. But we see some unusual estimator values when non-unit masses are considered, as shown in Figure \ref{fig_h+c}. From this figure we notice that for non-unit masses $\zeta^{(h+c)}$ is unusually low for NILC, SMICA and WMAP. The robustness of this signal is violated only for COMM which has a higher value of $\zeta^{(h+c)}$ compared to the other three cleaned maps. All the cleaned maps are representative of the same CMB signal, possibly barring some minor foreground residuals that differ among these maps. Such residuals could be causative of the differences in values of $\zeta^{(h+c)}$ between COMM and the other maps. We will study the cause of such differences in detail in a future work. In addition we find that the value of $\gamma^{(h+c)}$ is unusually high for NILC and WMAP, and low for SMICA (not shown in Figure \ref{fig_h+c}). Hence, this analysis illustrates how the use of non-unit masses is sensitive to any of the signals that could arise from the randomness of the peak values of hot spots and cold spots, as well as that from their eigenvector directions. This makes the use of non-unit masses in the orientation matrix a more general and inclusive approach. Hence, treating hot and cold spots separately with non-unit masses provides us with an opportunity to explore their distinct behaviours regarding isotropy.

\section{Summary and conclusion}\label{conc}

The principal property of isotropy of the distributions of hot spots and cold spots of the CMB is an important facet that needs to be studied by performing detailed investigations using foreground minimized CMB maps. An unusual observed property could give insights into new physics as regards the existence of any structure in the early universe. In this article, we presented a modified form of the orientation matrix to account for magnitudes of the data points on a 2-sphere. Eigenvalues of this matrix can be used to construct the so-called shape ($\gamma^{(s)}$) and strength ($\zeta^{(s)}$) parameters, where the superscript $s$ can be replaced with $h$ or $c$ to denote hot or cold spots, respectively. We employed these parameters as estimators to analyse distributions of hot and cold spots on the CMB maps at low resolution. The shape parameter helps distinguish clusters or girdles (rings) of the spots, and the strength parameter quantifies how strongly non-uniform (anisotropic) is their placement on the celestial sphere. Large scale homogeneity and isotropy can be investigated quantitatively with the help of these estimators. We demonstrated this with the help of two toy model CMB maps. The estimators were also evaluated for observed CMB maps and compared with those from pure CMB maps, which are Monte-Carlo simulations of statistically isotropic realisations of the CMB obtained using the $\Lambda CDM$ concordance model.

In our study we consider analysis over both full sky and partial sky CMB maps. For full sky analysis we use four foreground cleaned CMB maps, i.e., Commander (COMM), NILC and SMICA from Planck’s 2018 data-release, and WMAP’s 9 year ILC map (WMAP). For partial sky analysis we use several masks. These include non-Gaussian cold spot (NGCS) mask, WMAP $KQ75$ and Planck $U73$ masks. We also use two union masks which are determined by the pairs of masks $KQ75$, NGCS and $U73$, NGCS. For partial sky analysis with galactic masks, in addition to the four foreground cleaned maps mentioned above, we use WMAP’s foreground reduced Q, V, W frequency band maps (freqQ, freqV, freqW), and Planck’s foreground subtracted `fgsub-sevem' 70, 100 GHz maps (freq70, freq100) respectively, resulting in a set of nine foreground minimized CMB maps for this case. A summary of important observations stemming out from the work of this article is mentioned below.

Employing isotropy statistics over full sky we find
\begin{enumerate}
    \item[(i)] that the hot spots of all four cleaned maps exhibit highly uniform distributions. These correspond to consistently low values of $\zeta^{(h)}$ ($>95\%$ C.L.) when compared with pure CMB maps.
    \item[(ii)] The values of $\zeta^{(c)}$ are small. The distributions of cold spots are consistent with pure CMB realisations except for NILC and SMICA.
    \item[(iii)] Since the distributions of hot spots and that of cold spots tend to be uniform, neither type of spots for the cleaned maps show any signature of a ring or clustering nature.
    \item[(iv)]  Masking the NGCS only slightly increases the significance of low $\zeta^{(h)}$ for the four maps, while slightly reducing that of low $\zeta^{(c)}$ for SMICA and washing out the signal of low $\zeta^{(c)}$ for NILC. However, the conclusion of point (iii) above remains unchanged with respect to the NGCS mask.

\end{enumerate}

Analysing isotropy statistics over the $KQ75$ masked sky we find

\begin{enumerate}
    \item[(i)] that the signal of highly uniform placement of hot spots as seen for the full sky disappears. Instead, the cold spots for all nine maps tend to be very uniform. They are significantly uniform (at $> 95\%$ C.L.) for all the partial sky CMB maps except SMICA and freq100 maps.
    \item [(ii)] The significance of the above findings slightly increases after applying the NGCS mask.
    \item[(iii)] We do not find any clustering or girdling (ring) structure of either of hot or cold spots, as in the full-sky case.
\end{enumerate}

Instead of the $KQ75$ mask if we apply the $U73$ mask,
\begin{enumerate}
    \item[(i)] the distribution of cold spots becomes even more uniform, pushing the probability of observation of such high degree of uniformity compared to pure CMB realisations further into the critical region for all the partial sky CMB maps, except freq100.
    \item[(ii)] However, unlike the case of the KQ75 mask, when the NGCS mask is applied, the significance of the results slightly decrease for COMM, NILC, SMICA, freqV, and increase for WMAP, freqQ, freqW, and freq70.

    \item[(iii)] No specific signature of girdling or clustering is observed for hot or cold spots, except with freqQ for which some ring structure exists with and without the NGCS.

\end{enumerate}

Thus, with the analysis of partial sky outside galactic masks, very low $\zeta^{(c)}$ is found to be anomalous for most of the nine maps. This result is robust despite the use of two different masks (KQ75 and U73), data from different instruments (WMAP and Planck’s HFI and LFI) and at various frequencies, and masking of the NGCS. Exceptions to such robustness are SMICA and freq100 with the $KQ75$ mask and freq100 with the $U73$ mask. However the low values of $\zeta^{(c)}$ for these two maps are still close to being unlikely at $> 91\%$ C.L.

The strength of non-uniformity for hot spots ($\zeta^{(h)}$) is seen to be robustly low for all the cleaned maps (COMM, NILC, SMICA, and WMAP’s ILC) on the full sky and on the partial sky outside the NGCS mask. As for the partial sky analysis with galactic masks, we find that the low values of $\zeta^{(h)}$ are no longer significantly unexpected. This could mean that some galactic foreground residual common to all the four cleaned CMB maps is causative of the unusually low $\zeta^{(h)}$ when the galactic region is included for the analysis. Irrespective of its actual nature of origin, a washout of this signal is probable due to masking of the galactic region.

The use of galactic masks may not completely rule out any unknown foreground residuals or other systematic errors creeping in from the unmasked region of the sky. However, the simultaneous application of masks to both pure and observed CMB maps rules out a cut-sky effect as a causative agent of the anomalies observed.

The average peak values of hot spots and cold spots and the CMB temperature variance are known to be anomalously low. Since we consider peak values for studying non-uniformity of spots, therefore the signal of low strength of non-uniformity may be related with the low CMB temperature variance. The low variance anomaly is seen to vanish when the quadrupole and octupole are removed. Hence we perform our analysis again after using a cosine filter which excludes the quadrupole and octupole. A robust disappearance of the signal occurs across all cleaned maps, indicating that the low variance anomaly and the unusually weak non-uniformity of spots are potentially related. The robustness is lost on the partial sky outside galactic masks only for freqQ and freqV maps, which strongly suggests that some foreground residuals in these two maps contribute to the signal. 

The source of the signals observed in this article remains uncertain at present. Further investigation will be necessary for understanding the possible origin of the signals observed in this work. The robustness of the uniform signal of hot spots on the full sky and sudden spill over of the cold spot signal over most of the cleaned CMB maps (obtained from two different satellite missions, several different frequencies, detectors and foreground removal algorithms) excluding galactic regions, raises a significant curiosity as to whether the signals may be related to a cosmological origin. Additionally, both these signals are seen to be independent of the presence or absence of the non-Gaussian cold spot. However, we find that the signals of anomalously weak non-uniformity of spots could share a common origin with the low CMB temperature variance and anomalous contributions of the quadrupole and octupole.



\acknowledgements

We thank the anonymous referee for critical comments and suggestions that contributed to significant improvements in our manuscript. We acknowledge the use of the publicly available HEALPix \citep{2005ApJ...622..759G} software package (\url{http://healpix.sourceforge.io}). Our analyses are based on observations from Planck (\url{http://www.esa.int/Planck}), an ESA science mission with instruments and contributions directly funded by ESA Member States, NASA, and Canada. We acknowledge the use of the Legacy Archive for Microwave Background Data Analysis (LAMBDA), part of the High Energy Astrophysics Science Archive Center (HEASARC). HEASARC/LAMBDA is a service of the Astrophysics Science Division at the NASA Goddard Space Flight Center. 



\end{document}